\documentclass[12pt]{article}

\usepackage[a4paper,left=30mm,right=20mm,top=20mm,bottom=20mm]{geometry}
\usepackage{graphics}
\usepackage{epsfig}
\usepackage{cite}

\title{LHC soft physics and TMD gluon density at low $x$}

\author{A.V.~Lipatov$^{1,\,2}$, G.I.~Lykasov$^2$, N.P.~Zotov$^1$}

\begin{document}

\maketitle

\begin{center}

{\it $^1$Skobeltsyn Institute of Nuclear Physics, Lomonosov Moscow State University, 119991 Moscow, Russia}\\
{\it $^2$Joint Institute for Nuclear Research, Dubna 141980, Moscow Region, Russia}

\end{center}

\vspace{0.5cm}

\begin{center}

{\bf Abstract }

\end{center}

We study the unintegrated, or transverse momentum dependent (TMD)
gluon distribution obtained from the best description of the LHC data
on the inclusive spectra of hadrons produced in the mid-rapidity 
region and at low transverse momenta at the starting scale $Q_0^2 = 1$~GeV$^2$. 
To extend this gluon density to higher $Q^2$
we apply the Catani-Ciafoloni-Fiorani-Marchesini (CCFM) evolution equation.
The influence of the initial (starting) 
non-perturbative gluon distribution is studied.
The application of the obtained gluon density
to the analysis of the deep inelastic $ep$ scattering allows us to get the 
results which describe reasonably well the H1 and ZEUS data on the longitudinal 
proton structure function $F_L(x,Q^2)$. So, the connection
between the soft processes at the LHC and small $x$ physics at HERA has been confirmed
and extended to a wide kinematical region.

\vspace{1.0cm}

\noindent
PACS number(s): 12.38.Bx, 13.60.Hb

\newpage
\indent

Usually, the scale-dependent parton density distribution is  
calculated as a function of the Bjorken variable $x$
and the square of the four-momentum transfer $q^2 = - Q^2$
within the framework of the Dokshitzer-Gribov-Lipatov-Altarelli-Parisi 
(DGLAP) evolution equations\cite{1} based on standard 
collinear QCD factorization. However, for semi-inclusive processes
(such as inclusive jet production in $ep$ deep inelastic scattering (DIS), heavy flavour production in hadron collisions etc.) 
at high energies, which are sensitive to the details of 
parton kinematics, it is more appropriate
to use the parton distributions unintegrated over the partonic transverse momentum $k_T$, or 
transverse momentum dependent (TMD) distributions\cite{2}.
The latter are a subject of intense studies, and various approaches to the
investigation of these quantities have been proposed\cite{3,4,5,6}.
Recently, two basic TMD gluon densities have been used in the
small-$x$ formalism, the so-called Weizsaker-Williams gluon distribution and
the dipole one\cite{7,8,9}. In general, at asymptotically large energies (or very small $x$) the 
theoretically correct description of TMD gluon densities is based on the 
Balitsky-Fadin-Kuraev-Lipatov (BFKL) evolution equation\cite{10} 
where the leading $\ln(1/x)$ contributions are 
taken into account in all orders. 
Another approach, which is valid for both small and large $x$, is given by the CCFM  
gluon evolution equation\cite{11}. It introduces angular ordering of emissions to correctly treat the 
gluon coherence effects. In the limit of asymptotically high energies, it is almost equivalent to BFKL 
and also similar to the DGLAP evolution for large $x \sim 1$. The resulting TMD gluon 
density depends on two scales; the additional scale $\bar q^2$ is a variable 
related to the maximum angle allowed in the emission and plays the role of the evolution 
scale $\mu^2$ in the collinear parton densities.
Early phenomenological applications
of TMD partons within the framework of the $k_T$-factorization QCD approach\cite{12,13} 
can be found \cite{2}.

In the present note we concentrate mostly on the TMD gluon density proposed in\cite{14}.
This gluon density was calculated within the soft QCD model as a function of $x$ and ${\mathbf k}_T^2$ at a fixed 
value of the scale $Q_0^2 = 1$~GeV$^2$ and can be presented in the simple analytical form
\begin{equation}
\displaystyle f_g^{(0)}(x,{\mathbf k}_T^2,Q_0^2) = \frac{3\sigma_0}{4\pi^2\alpha_s} C_1 (1-x)^{b_g} \times \atop {
\displaystyle \times \left[R_0^2(x){\mathbf k}_T^2 + C_2\left(R_0^2(x){\mathbf k}_T^2\right)^{a/2}\right] \exp\left(-\left[R_0^2(x)
{\mathbf k}_T^2\right]^{1/2}-d\left[R_0^2(x){\mathbf k}_T^2\right]^{3/2}\right)},
\end{equation}

\noindent 
where $R_0^2(x) = (x/x_0)^\lambda/Q_0^2$ and all parameters $\sigma_0 = 29.12$ mb, $C_1 = 0.3295$, $C_2 = 2.3$, 
$a = 0.7$, $b_g = 12$, $d = 0.2$, $x_0 = 4.1 \cdot 10^{-5}$, $\lambda = 0.22$ and $\alpha_s = 0.2$ 
are found from the best fit of the LHC data on the inclusive spectrum of charged
hadrons produced in $pp$ collisions in the mid-rapidity region at small $p_T\leq$1.6 GeV\cite{14}.
The $q\bar q$ dipole cross section, derived from the proposed gluon density 
as a function of the transverse distance $r$ between
$q$ and $\bar q$ in the dipole, differs from the one calculated in\cite{15,16,17,18,19}. In particular, 
it is saturated earlier with increasing $r$ more
than the dipole cross section predicted by the Golec-Biernat-Wusthoff (GBW) saturation model\cite{18}. 
It is connected with the $x$ dependence of the gluon distribution~(1),
which is different from the GBW gluon\cite{18} at small intrinsic transverse momenta 
$|{\mathbf k}_T| < 1$ or $1.5$~GeV and coincides with it at larger $|{\mathbf k}_T| > 1.5$~GeV
at fixed $Q_0^2 = 1$~GeV$^2$. 

The gluon density~(1) was used to calculate the proton structure 
functions $F_2^{c}(x,Q^2)$, $F_2^{b}(x,Q^2)$ and $F_L(x,Q^2)$, and a reasonably good
description of the H1 and ZEUS data at low and moderate $Q^2$ was obtained\cite{14}.
On this basis, the connection between the soft processes at LHC and 
small $x$ physics at HERA was claimed.
The question arises, what will be at any $Q^2$ and how the observables
like the proton structure functions will change. Actually, this is the main subject of our paper.
Below we continue the previous analysis\cite{14} and extend the consideration to the whole kinematical range.
We will treat the proposed gluon density as an initial (starting) distribution and 
use the CCFM evolution equation which is the most natural tool to study details of  
the perturbative and non-perturbative QCD evolution\footnote{See \cite{2} 
for more information.}. This equation with respect to the evolution (factorization) 
scale $\bar q^2$ can be written as \cite{11}
\begin{equation}
\displaystyle f_g(x,{\mathbf k}_T^2,\bar q^2) = f_g^{(0)}(x,{\mathbf k}_T^2,Q_0^2) \Delta_s(\bar q^2,Q_0^2) + \atop { 
\displaystyle + \int {dz\over z} \int {d q^2\over q^2} \theta(\bar q - z q) \Delta_s(\bar q^2,q^2)} P_{gg}(z,q^2,{\mathbf k}_T^2) 
f_g(x/z,{\mathbf k^\prime}_T^2,q^2),
\end{equation}

\noindent 
where ${\mathbf k^\prime}_T = {\mathbf q} (1 - z)/z + {\mathbf k}_T$ and the 
Sudakov form factor $\Delta_s(q_1^2,q_2^2)$ describes the probability of no radiation
between $q_2^2$ and $q_1^2$. 
The first term in the CCFM equation gives the contribution of
non-resolvable branchings between the starting scale $Q_0^2$ and the factorization scale
$\bar q^2$, the second term describes the details of QCD evolution expressed by the 
convolution of the CCFM splitting function $P_{gg}(z,q^2,{\mathbf k}_T^2)$ with the
gluon density $f_g(x,{\mathbf k}_T^2,\bar q^2)$ and the Sudakov form factor $\Delta_s(\bar q^2,q^2)$, and the theta function introduces 
angular ordering
of emissions to correctly treat the gluon coherence effects.
The evolution scale $\bar q^2$ is defined by the maximum allowed angle for any emission.
The analytical expressions for the splitting function $P_{gg}(z,q^2,{\mathbf k}_T^2)$
and the Sudakov form factor can be found in\cite{11}.

The CCFM evolution equation~(2) with the starting distribution~(1) was 
solved numerically\footnote{Authors are very grateful to Hannes Jung for providing 
us with the appropriate numerical code.} 
using the Monte Carlo method, and the resulting TMD gluon density was 
obtained for any values of $x$, ${\mathbf k}_T^2$ and the hard scale $\mu^2$
(below we would not distinguish $\mu^2$ and $\bar q^2$).
The corresponding data file is available from the authors upon request\footnote{lipatov@theory.sinp.msu.ru}.
In Fig.~1 the calculated gluon density $f_g(x,{\mathbf k}_T^2,\mu^2)$ is shown as a function of ${\mathbf k}_T^2$
for different values of $\mu^2$ at fixed $x = 10^{-4}$.
The contributions from the first term in the CCFM evolution equation (i.e. starting gluon distribution)
are shown separately. One can see that the influence of this initial distribution is concentrated
at small values of ${\mathbf k}_T^2$, whereas at ${\mathbf k}_T^2 > 1$~GeV$^2$
the perturbative evolution is important.
Moreover, at ${\mathbf k}_T^2 \geq 10$~GeV$^2$ the CCFM evolution results in the increase 
of the gluon density by a few orders of magnitude. So, the inclusion of the CCFM evolution for 
$f_g(x,{\mathbf k}_T^2,\mu^2)$ is very important at low $x$ and $|{\mathbf k}_T|$ above a few~GeV,
especially at large values of $\mu^2$.  
To illustrate the non-perturbative effects connected with the
small ${\mathbf k}_T^2$ region we replace the initial gluon density~(1)
by the GBW gluon distribution\cite{18} derived from the popular 
GBW saturation model and repeat the CCFM evolution procedure
in the same manner as described above.
The resulting gluon density and the pure GBW gluon distribution
are also shown in Fig.~1.
Note that here the dashed curves practically coincide with the solid ones at ${\mathbf k}_T^2 < 1$~GeV$^2$ 
and at ${\mathbf k}_T^2 > 1$~GeV$^2$ they coincide with the dotted curves. 
Also, the dash-dotted curves
are very similar to the dotted curves at small ${\mathbf k}_T^2$ and to the 
solid ones at large ${\mathbf k}_T^2$.
Note that even with very different starting distributions,
the TMD gluon densities after perturbative CCFM evolution are similar at large ${\mathbf k}_T^2$.
Therefore, the small ${\mathbf k}_T^2$ region provides information on the non-perturbative 
part of the parton density functions.

As was mentioned above, the TMD gluon density given by~(1) was used \cite{14} in 
the analysis of the recent HERA data on the proton structure functions $F_2^{c}(x,Q^2)$, 
$F_2^{b}(x,Q^2)$ and $F_L(x,Q^2)$. Below we use the obtained CCFM-evolved gluon 
distribution to describe the H1 and ZEUS data\cite{20,21} on the longitudinal structure function
$F_L(x,Q^2)$, which is directly connected to the gluon content of the
proton.
It is equal to zero in the parton model with spin $1/2$ partons and has nonzero values within the 
 pQCD\footnote{We do not consider here charm and beauty contributions to 
the proton structure function $F_2(x,Q^2)$ due to lack of space.}.
The consideration is based on main formulas which have been listed in\cite{22}. Here we
only recall some of them. According to the $k_T$-factorization prescription, 
the proton structure function $F_L(x,Q^2)$ can be calculated
as
\begin{equation}
F_L(x, Q^2) = \sum_f e_f^2 \int {dy \over y} \int d{\mathbf k}_T^2 \, {\cal C}_L(x/y,{\mathbf k}_T^2,Q^2,m_{f}^2,\mu^2) 
f_g(y,{\mathbf k}_T^2,\mu^2),
\end{equation}

\noindent 
where $f$ is the quark flavour, and $e_f$ and $m_f$ are the electric charge and mass of the quark. 
The hard coefficient function ${\cal C}_{L}(x,{\mathbf k}_T^2,Q^2,m^2,\mu^2)$ corresponds to the quark-box diagram 
for the photon-gluon fusion subprocess and was calculated in \cite{22}.
Numerically, we set the masses of the charm and beauty quarks to $m_c = 1.4$~GeV and $m_b = 4.75$~GeV
and use the massless limit to evaluate the corresponding contributions from the light quarks.
Also, we apply the LO formula for the strong coupling constant $\alpha_s(\mu^2)$ with $n_f = 4$ quark flavours
at $\Lambda_{\rm QCD} = 200$~MeV, such that $\alpha_s(M_Z^2) = 0.1232$.
Note that in all aspects we strictly follow our previous consideration\cite{14,22}.
In order to take into account the NLO corrections (which are important at low $Q^2$),
we use the shifted value of the 
renormalization scale $\mu_R^2 = K\,Q^2$, where $K\sim 127$. As
shown in\cite{23}, this shifted scale in the DGLAP approach at the LO approximation leads to the
results which are very close to the NLO predictions. In the $k_T$-factorization approach,
this procedure gives us a possibility of taking into account additional higher-twist and
non-logarithmic NLO corrections \cite{22}.

The results of our calculations are presented in Figs.~2 and~3. We
show separately the predictions obtained with the 
proposed CCFM-evolved gluon distribution and 
the results of the calculations based on the pure starting gluon density~(1).
One can see that the predictions obtained with the proposed CCFM-evolved gluon density agree well with the H1 
and ZEUS data in the whole kinematical region of $x$ and $Q^2$, whereas the 
non-evolved distribution~(1) fits well the data at small $Q^2$ only
and tends to underestimate them at large $Q^2$.
We find that the inclusion of the CCFM evolution is very important and has important consequences, 
both qualitative and quantitative. In particular, it changes the shape 
of the calculated longitudinal structure function $F_L(x,Q^2)$, especially at low $x$.
Therefore we conclude that the link between soft processes at the LHC and low-$x$
physics at HERA, pointed out in\cite{14} for small $Q^2$, 
is confirmed and extended now to a wide kinematical region.
Additionally, we show the results obtained with the 
TMD gluon distribution where the GBW gluon density is used 
as an input for the CCFM evolution as described above. One can see that the influence of
the shape and other parameters of the initial non-perturbative gluon distribution on the 
description of the collider data is significant for a wide
region of $x$ and $Q^2$.
The proposed TMD gluon density where all these parameters are
verified by the description of the LHC data on the hadron spectra 
in the soft kinematical region leads to the best agreement with the HERA data
according to Table~1, where the corresponding $\chi^2$/d.o.f. values are presented.
It is important for further phenomenological investigations
of small-$x$ physics at the LHC.

\begin{table}
\begin{center}
\begin{tabular}{|l|c|c|}
\hline
   & & \\
  Source & $\chi^2$/d.f. (H1) & $\chi^2$/d.o.f. (ZEUS)\\
   & & \\
\hline
   & & \\
   CCFM-evolved gluon density~(1) & $7.20/12$ & $2.219/5$ \\
   & & \\
   CCFM-evolved GBW gluon density & $5.39/12$ & $0.68/5$\\
   & & \\
\hline
\end{tabular}
\end{center}
\caption{The estimated $\chi^2$/d.o.f. values for our fit of the H1\cite{20} and ZEUS\cite{21} data.}
\label{table1}
\end{table}

{\sl Acknowledgements.}
We thank H.~Jung for his big help in the calculation of the CCFM evolution for the
TMD gluon distribution and useful discussions.
The authors are also grateful to B.I.~Ermolaev, K.~Kutak and D.~Toton 
for discussions and comments. A.V.L. and N.P.Z. are very grateful to the
DESY Directorate for the support within the Moscow --- DESY project on Monte-Carlo
implementation for HERA --- LHC.
This research was supported in part by the FASI of the Russian Federation
(grant NS-3920.2012.2), RFBR grants 12-02-31030, 11-02-01538-a and 13-02-01060 and the grant of the Ministry of education and sciences
of Russia (agreement 8412).

\newpage

\begin{figure}
\begin{center}
\epsfig{figure=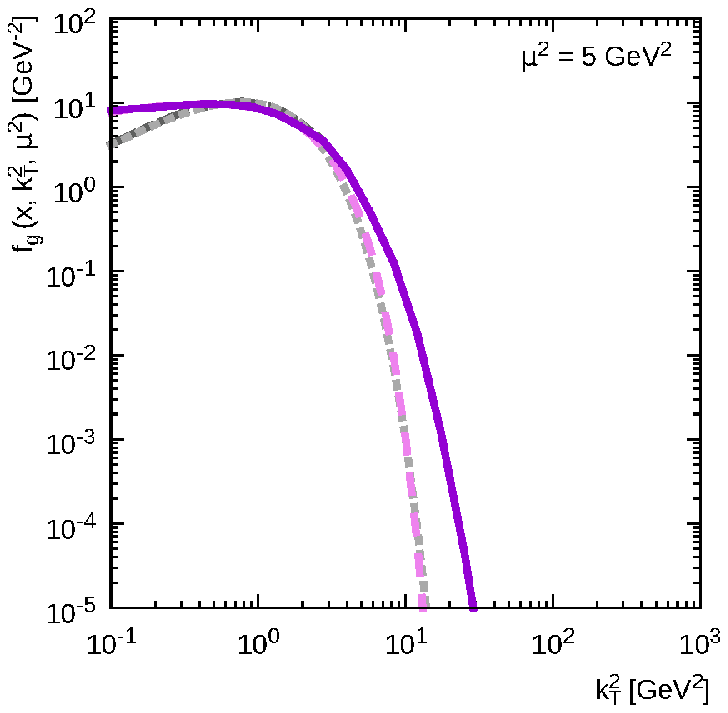, width = 7.9cm}
\epsfig{figure=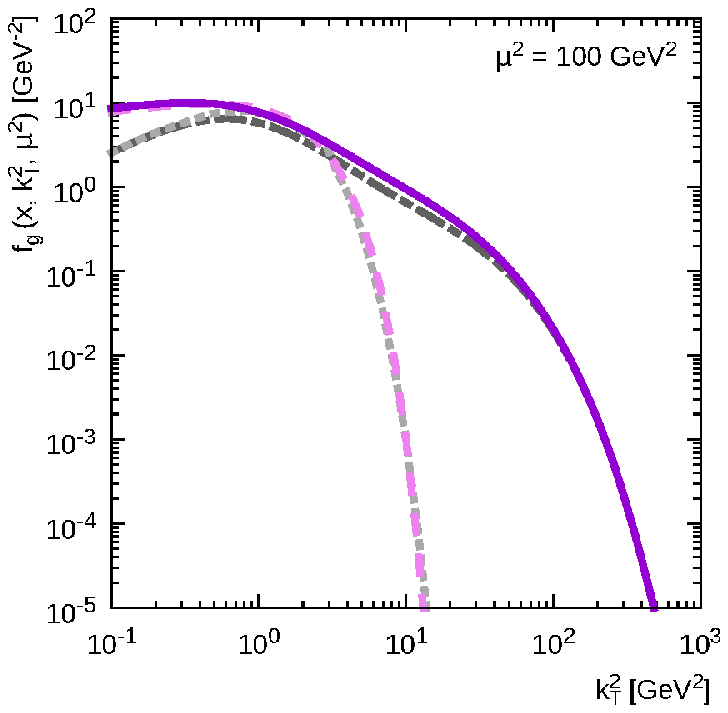, width = 7.9cm}
\caption{Comparison of the different TMD gluon densities 
as a function of ${\mathbf k}_T^2$ for $\mu^2 = 5$~GeV$^2$ (left panel) and $\mu^2 = 100$~GeV$^2$ 
(right panel) at fixed $x = 10^{-4}$.
The solid curves correspond to the proposed CCFM-evolved gluon density.
The contributions from initial gluon distribution~(1) 
are shown by the dashed curves. The dash-dotted and dotted curves
correspond to the CCFM-evolved GBW gluon density and the pure (non-evolved) GBW gluon, respectively. }
\label{fig1}
\end{center}
\end{figure}

\begin{figure}
\begin{center}
\epsfig{figure=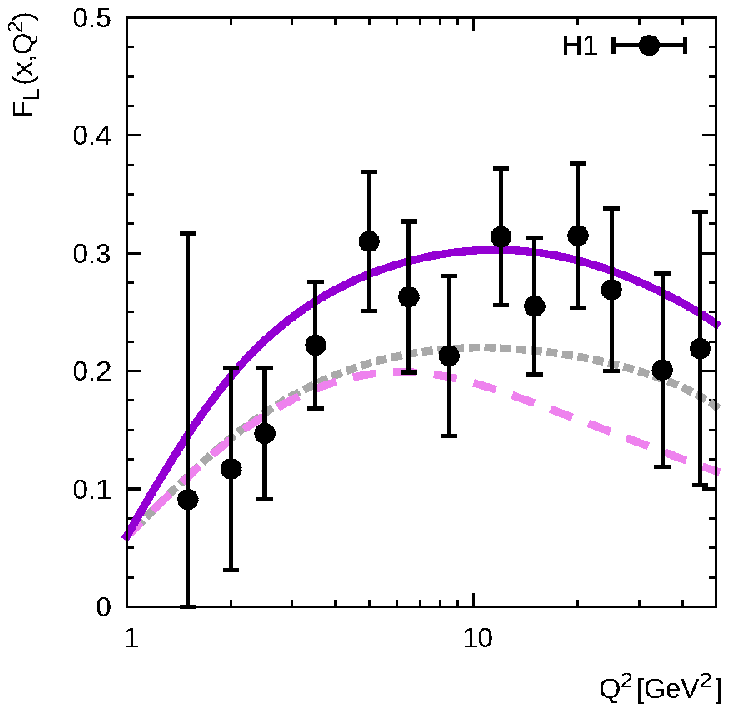, width = 7.9cm}
\epsfig{figure=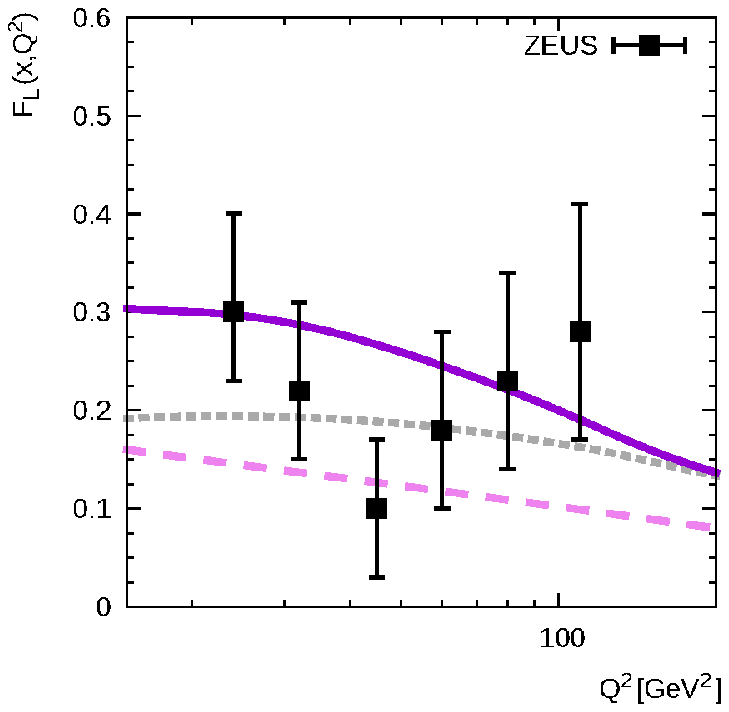, width = 7.9cm}
\caption{The longitudinal proton structure function $F_L(x,Q^2)$ 
as a function of $Q^2$. The solid curves correspond to the results obtained with the 
proposed CCFM-evolved TMD gluon density, and 
the contributions from initial gluon distribution given by~(1) 
are shown by the dashed curves. The dotted curves
correspond to the results obtained with the CCFM-evolved GBW gluon density.
The experimental data are from H1\cite{20} and ZEUS\cite{21}. 
In the ZEUS measurements the ratio $Q^2/x$ is a constant for each bin, which corresponds to
$y = 0.71$ and $\sqrt s = 225$~GeV, where $y = Q^2/xs$.}
\label{fig2}
\end{center}
\end{figure}

\newpage

\begin{figure}
\begin{center}
\epsfig{figure=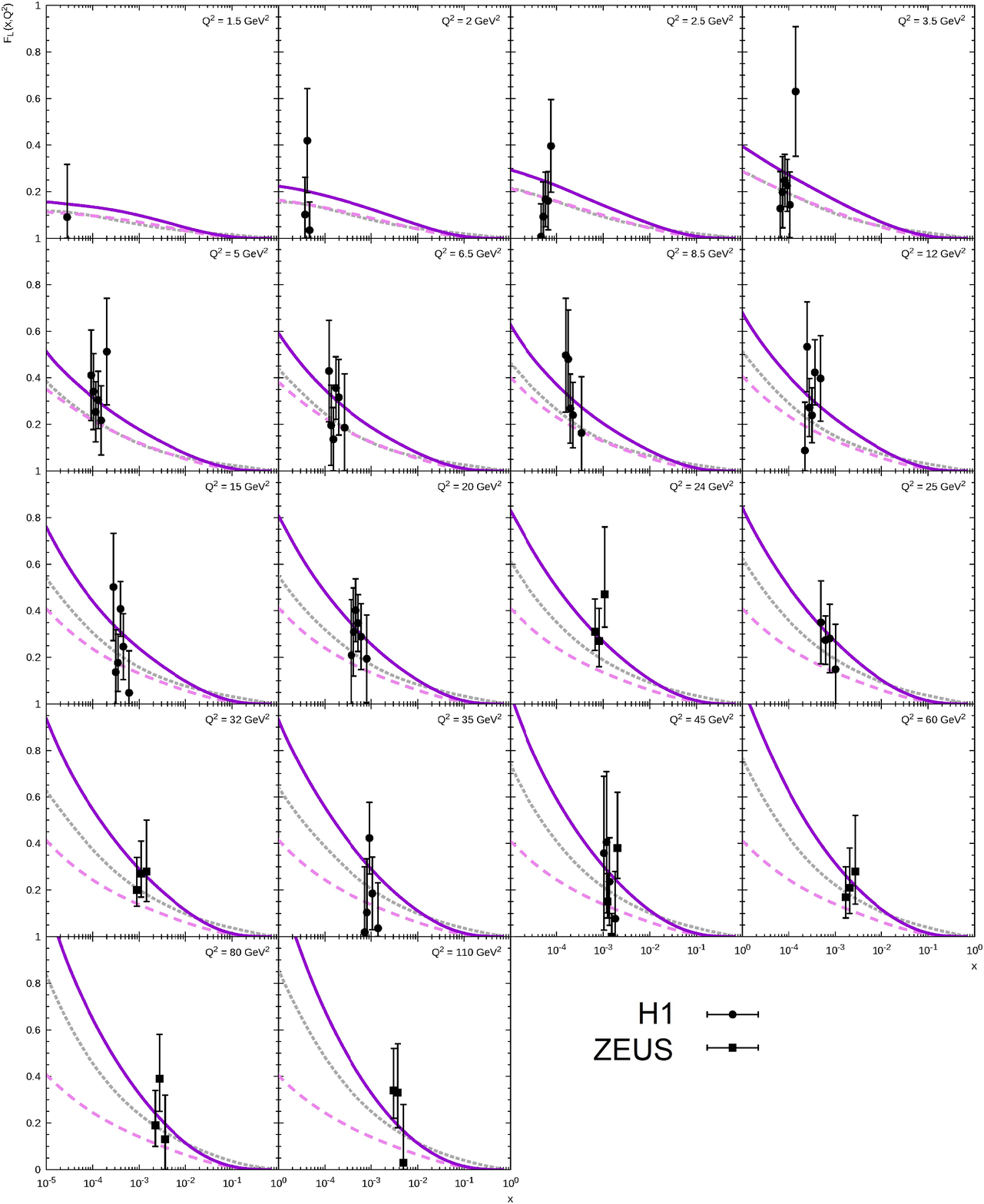, width = 16cm}
\caption{The longitudinal proton structure function $F_L(x,Q^2)$ 
as a function of $x$. Notation of all curves is the same as in Fig.~2.
The experimental data are from H1\cite{20} and ZEUS\cite{21}.}
\label{fig3}
\end{center}
\end{figure}


\begin{thebibliography}{99}
\bibitem{1} V.N.~Gribov and L.N.~Lipatov, Sov.J. Nucl. Phys. {\bf 15}, 438 (1972);\\
  L.N.~Lipatov, Sov. J. Nucl. Phys.{\bf 20}, 94 (1975);\\
  G.~Altarelli, G.~Parisi, Nucl. Phys. B {\bf 126}, 298 (1977);\\
  Yu.L.~Dokshitzer, Sov. Phys. JETP, {\bf 46}, 641 (1977).
\bibitem{2} B.~Andersson {\sl et al.} (Small-$x$ Collaboration), Eur. Phys. J. C {\bf 25}, 77 (2002);\\
  J.~Andersen {\sl et al.} (Small-$x$ Collaboration), Eur. Phys. J. C {\bf 35}, 67 (2004);\\
  J.~Andersen {\sl et al.} (Small-$x$ Collaboration), Eur. Phys. J. C {\bf 48}, 53 (2006).
\bibitem{3} J.C.~Collins, {\it Foundations of perturbative QCD}, Cambridge University Press, 2011.
\bibitem{4} E.~Avsar, arXiv:1108.1181 [hep-ph]; arXiv:1203.1916 [hep-ph].
\bibitem{5} F.~Dominguez, C.~Marquet, B.-W.~Xiao, F.~Yuan, Phys. Rev. D {\bf 83}, 105005 (2011).
\bibitem{6} S.M.~Aybat, T.C.~Rogers, Phys. Rev. D {\bf 83}, 114042 (2011).
\bibitem{7} J.-W.~Qiu, P.~Sun, B.-W.~Xiao, F.~Yuan, arXiv:1310.2230 [hep-ph].
\bibitem{8} P.~Sun, C.-P.~Yuan, F.~Yuan, Phys. Rev. D {\bf 88}, 054008 (2013).
\bibitem{9} A.H.~Mueller, B.-W.~Xiao, F.~Yuan, arXiv:1308.2993 [hep-ph].
\bibitem{10} E.A.~Kuraev, L.N.~Lipatov, V.S.~Fadin, Sov. Phys. JETP {\bf 44}, 443 (1976);\\
  E.A.~Kuraev, L.N.~Lipatov, V.S.~Fadin, Sov. Phys. JETP {\bf 45}, 199 (1977);\\
  I.I.~Balitsky, L.N.~Lipatov, Sov. J. Nucl. Phys. {\bf 28}, 822 (1978).  
\bibitem{11} M.~Ciafaloni, Nucl. Phys. B {\bf 296}, 49 (1988);\\
  S.~Catani, F.~Fiorani, G.~Marchesini, Phys. Lett. B {\bf 234}, 339 (1990);\\
  S.~Catani, F.~Fiorani, G.~Marchesini, Nucl. Phys. B {\bf 336}, 18 (1990);\\
  G.~Marchesini, Nucl. Phys. B {\bf 445}, 49 (1995). 
\bibitem{12} L.V.~Gribov, E.M.~Levin, M.G.~Ryskin, Phys. Rep. {\bf 100}, 1 (1983);\\
  E.M.~Levin, M.G.~Ryskin, Yu.M.~Shabelsky, A.G.~Shuvaev, Sov. J. Nucl. Phys. {\bf 53}, 657 (1991).
\bibitem{13} S.~Catani, M.~Ciafoloni, F.~Hautmann, Nucl. Phys. B {\bf 366}, 135 (1991);\\
  J.C.~Collins, R.K.~Ellis, Nucl. Phys. B {\bf 360}, 3 (1991).
\bibitem{14} A.A.~Grinyuk, A.V.~Lipatov, G.I.~Lykasov, N.P.~Zotov, Phys. Rev. D {\bf 87}, 074017 (2013).
\bibitem{15} N.N.~Nikolaev, B.G.~Zakharov, Z. Phys. C {\bf 49}, 607 (1991).
\bibitem{16} I.P.~Ivanov, N.N.~Nikolaev, Phys. Rev. D {\bf 65}, 054004 (2002).
\bibitem{17} J.~Nemchik, V.~Barone, M.~Genovese, N.N.~Nikolaev, E.~Predazzi,
  B.G.~Zakharov, Phys. Lett. B {\bf 326}, 161 (1994).
\bibitem{18} K.~Golec-Biernat, M.~Wusthoff, Phys. Rev. D {\bf 59}, 014017 (1998);\\
  K.~Golec-Biernat, M.~Wusthoff, Phys. Rev. D {\bf 60}, 114023 (1999).
\bibitem{19} J.L.~Albacete, C.~Marquet, Phys. Lett. B {\bf 687}, 174 (2010).
\bibitem{20} F.D.~Aaron {\sl et al.} (H1 Collaboration), Eur. Phys. J. C {\bf 71}, 1579 (2011).
\bibitem{21} S.~Chekanov {\sl et al.} (ZEUS Collaboration), Phys. Lett. B {\bf 682}, 8 (2009).
\bibitem{22} A.V.~Kotikov, A.V.~Lipatov, G.~Parente, N.P.~Zotov, Eur. Phys. J. C {\bf 26}, 51 (2002);\\
  A.V.~Kotikov, A.V.~Lipatov, N.P.~Zotov, Eur. Phys. J. C {\bf 27}, 219 (2003);\\
  A.V.~Kotikov, A.V.~Lipatov, N.P.~Zotov, JETP {\bf 101}, 811 (2005).
\bibitem{23} S.J.~Brodsky, V.S.~Fadin, V.T.~Kim, L.N.~Lipatov, G.B.~Pivovarov, JETP Lett. {\bf 70}, 155 (1999).

\end{thebibliography}
\end{document}